%% file: negutrap.tex
\documentclass[aps,prb,twocolumn,superscriptaddress,showpacs]{revtex4-1}

\usepackage{graphicx}
\usepackage{color}
\usepackage{xspace}
\usepackage{amsmath}
\usepackage[suffix=]{epstopdf}
\usepackage{nicefrac}

\include{definitions}

\begin{document}

\title{Superconductivity and charge order of confined Fermi systems}

\date{\today}

\author{E. Assmann}
\email{assmann@ifp.tuwien.ac.at}
\affiliation{Institut f\"ur Theoretische Physik, Technische
  Universit\"at Graz, Petergasse 16, 8010 Graz, Austria}
\affiliation{Institut f\"ur Festk\"orperphysik, Technische
  Universit\"at Wien, Wiedner Hauptstraße 8-10, 1040 Wien, Austria}

\author{S. Chiesa}
\affiliation{Physics Department, University of California, Davis,
  California 95616, USA}
\affiliation{Department of Physics and Astronomy, The University of
  Tennessee, 1408 Circle Dr., Knoxville, TN 37996-1200}

\author{G.G. Batrouni}
\affiliation{INLN, Universit\'e de Nice-Sophia Antipolis, CNRS; 1361
  route des Lucioles, 06560 Valbonne, France}
\affiliation{Institut Universitaire de France}

\author{H.G. Evertz}
\affiliation{Institut f\"ur Theoretische Physik, Technische
  Universit\"at Graz, Petergasse 16, 8010 Graz, Austria}

\author{R.T. Scalettar}
\affiliation{Physics Department, University of California, Davis,
  California 95616, USA}

\begin{abstract}
  The low-temperature properties of the two-dimensional attractive
  Hubbard model are strongly influenced by the fermion density.  Away
  from half-filling, there is a finite-temperature transition to a
  phase with $s$-wave pairing order.  However, $\tcrit$ is suppressed
  to zero at half-filling, where long-range charge-density-wave order
  also appears, degenerate with superconductivity.  This paper
  presents Determinant Quantum Monte Carlo simulations of the
  attractive Hubbard model in the presence of a confining potential
  $\vtrap$ which makes the fermion density $\density$ inhomogeneous
  across the lattice.  Pair correlations are shown to be large at low
  temperatures in regions of the trapped system with incommensurate
  filling, and to exhibit a minimum as the local density
  $\density(\vec i)$ passes through one fermion per site.  In this
  ring of $\density(\vec i)=1$, charge order is enhanced.  A
  comparison is made between treating $\vtrap$ within the
  local-density approximation (LDA) and in an \emph{ab initio} manner.
  It is argued that certain sharp features of the LDA result at
  integer filling do not survive the proximity of doped sites.  The
  critical temperature of confined systems of fixed characteristic
  density is estimated.
\end{abstract}


\maketitle

\section*{Introduction}

Studies of the interplay of spatial inhomogeneity and
superconductivity have a long history.  A seminal early result was
Anderson's realization \cite{anderson} that although the breaking of
translation invariance by disorder renders momentum no longer a good
quantum number, pairing still occurs between appropriately chosen
(time reversed) states.  Numerical studies within the Bogoliubov-de
Gennes approximation, \cite{ghosal98,aryanpour07} Quantum
Monte Carlo (QMC) \cite{mondaini08}, and other
approaches 
have quantified the magnitude of disorder which superconductivity can
withstand \cite{hebard94}.  In these studies, and the granular superconducting
materials they model \cite{orr86}, regions of pairing order coexist
with normal, or insulating, phases.  Superconductivity can be
destroyed by various mechanisms, including phase fluctuations between
the order parameter on different islands where Cooper pairs of bosons
exist,\cite{dynes84} or breaking of the Cooper pairs
themselves.\cite{dynes86} The dominant mechanism determines the
appropriate modeling, e.g.~a description within the disordered boson
\cite{fisher89} or fermion Hubbard Hamiltonians, or ``phase-only''
descriptions with the XY model and its variants.\cite{cha91}

Recently, experiments on ultra-cold atoms have provided a rather
different realization of inhomogeneity in the form of a smoothly
varying confining potential which produces a system with a radial
density profile, maximal at the trap center and falling off to zero at
the periphery.\cite{footnote1} Much attention has focussed on
repulsively interacting bosons and fermions.\cite{stoferle06,chin06}
In this case, a Mott insulator may coexist with superfluid or normal
phases.  For fermions, the Mott insulator also exhibits
antiferromagnetic correlations at very low temperatures.  At present,
experimentally accessible temperatures for fermionic systems are such
that a degenerate Fermi gas has been observed\cite{kohl05}, along with
signatures of the Mott phase.\cite{jordens08,schneider08} The ultimate
objective is insight into the ground state physics of the repulsive
Hubbard model (RHM), and, in particular, the fundamental issue of $d$-wave
superconducting order and its interplay with
antiferromagnetism.\cite{scalapinoxx}

This goal for repulsive fermions awaits the attainment of lower
experimental temperatures.  In the interim, it is useful to perform
careful studies of attractive systems.  This case is not only of
interest in its own right, but also QMC simulations can often attain
lower temperatures for attractive models, and thus can track
experiments closer to transitions into ordered
phases.

The focus of the present paper is the description of the behavior of
attractively interacting fermions in a two-dimensional confining
potential.  Some of the issues are similar to the repulsive case, in
particular the coexistence of phases as the density varies across the
trap.\cite{varney09} However, the attractive case has several
important distinctions, specifically the existence of known
finite-temperature phase transitions in two dimensions.  In addition,
in the repulsive case there is a broad range of chemical potentials
$\mu$ which fall within the ``Mott gap'' and for which the fermion
density $\density=1$.  That is, the compressibility
$\kappa=\partial\density/\partial \mu = 0$ at $\density=1$.  For the
confined system, this implies an extended region of commensurate
density, spatial sites which have a value of the local confining
potential which falls within the Mott gap.  In the attractive case,
the compressibility is finite ($\kappa \neq 0$) at commensurate
density.  As a consequence, the region of half-filling is a truly
one-dimensional ring as opposed to an annulus of finite thickness.

A key result of this work is that the unique features of charge
density wave physics at the single value of chemical potential which
gives commensurate filling do not survive coupling to neighbors of
incommensurate density.  Thus the correlations which appear in a
homogeneous system with commensurate filling are never achieved in a
trap; the Local-Density Approximation (LDA), in which the behavior of
each site in a confining potential is assumed to be that of a
homogeneous system with global density matching the local filling,
breaks down at that point.

This paper is organized as follows: In the next section we describe
the specific Hamiltonian, the attractive Hubbard model (AHM) and
aspects of the computational methodology, Determinant Quantum Monte
Carlo (DQMC), which will be used.  Results are then presented within
the LDA as well as from direct simulations of confined systems, and
the two approaches are compared.  Next, we present a finite-size
extrapolation using data from systems of different sizes at constant
characteristic density, and estimate the critical temperature of the
confined AHM.  A concluding section summarizes the results and
indicates some remaining open questions.

Studies of the AHM with inhomogeneity have been performed with
Variational Monte Carlo,\cite{fujihara07} Bogoliubov
de-Gennes,\cite{chen07,trivedi98} and Gutzwiller
approaches.\cite{ruegg07,yamashita07} Of particular relevance here is
work within dynamical mean field theory (DMFT) and a two-site impurity
solver \cite{koga09}, which suggested that the half-filled physics is
stabilized by a confining potential, and that an extended supersolid
phase of commensurate density exists in a trap.

\section*{Models and Computational Approach}

The attractive Hubbard Hamiltonian, in the presence of a 
confining potential, reads
\begin{multline}
  \label{eq:hamiltonian}
  \hat H = -t \sum_{\langle \vec i\vec j \rangle}
  ( c^{\dagger}_{\vec{i}\sigma} c^{}_{\vec{j}\sigma}
  + c^{\dagger}_{\vec{j}\sigma} c^{}_{\vec{i}\sigma} )
  - |U| \sum_{\vec i} ( n_{\vec{i}\uparrow} - \nicefrac12 )
  ( n_{\vec{i}\downarrow} - \nicefrac12 )
  \\
  - \sum_{\vec i} \left\{
  \mu - \vtrap\;  \dist^2
  \right\} (n_{\vec{i}\uparrow} + n_{\vec{i}\downarrow})\,.
\end{multline}
Here $c^{\dagger}_{\vec{i}\sigma}\: ( c^{}_{\vec{i}\sigma} )$ are
creation (destruction) operators at spatial site $\vec{i}$ for two
different species of fermions $\sigma$.  We choose the center of the
trap to be at a plaquette center and set the origin there, so that the
coordinates $i_x$ and $i_y$ take half-integer values.  In the
condensed-matter context, $\sigma=\pm\frac12$ is the electron spin.
For cold atoms, $\sigma$ labels two hyperfine states.  We will
consider the case of square lattices of linear size $L$.  The hopping
parameter $t$ can be tuned by changing the optical lattice
depth,\cite{footnote2}; in the following, $t=1$ is chosen to set the
scale of energy.  The sum $\sum_{\langle \vec i \vec j \rangle}$ is
over all near-neighbor pairs of sites, and $\sum_{\vec i}$ is over all
sites.  The on-site attraction $|U|$ can be tuned through the
application of a magnetic field via a Feshbach resonance.  The
chemical potential $\mu$ is set to get the desired number of particles
$\npart$.  Finally, $\vtrap$ is the trap curvature which determines
the strength of the confining potential.

In DQMC\cite{blankenbecler81}, the partition function $Z = \trace
e^{-\beta \hat H}$ is written as a path integral by discretizing the
inverse temperature $\beta$ into $M$ intervals of size $\Delta \tau =
\beta/M$.  The Trotter approximation \cite{trotter} $e^{-\Delta \tau
  \hat H} \simeq e^{-\Delta \tau \hat K} e^{-\Delta \tau \hat V} $
isolates the quartic terms (involving the interaction $U$) in $\hat
H$, and a discrete Hubbard-Stratonovich field \cite{hirsch85}
decouples $e^{-\Delta \tau \hat V}$ so that only quadratic terms in
the fermion operators appear.  When the trace over fermion operators
is done, $Z$ is expressed as a sum over the different field
configurations with a weight which is the product of two determinants
(one for each value of $\sigma$) of matrices with dimension $L^2\times
L^2$ given by the number of lattice sites.  In the case of attractive
$U$, because the two species couple to the Hubbard-Stratonovich field
with the same sign, the two determinants are identical and there is no
sign problem.\cite{loh90} This allows us to study confined systems
down to arbitrarily low temperatures, unlike the repulsive model where
the largest $\beta$ accessible is $\beta \simeq 3 \text{--} 4$ for
confined systems with $U = 4 \text{--} 8$.\cite{varney09}

The observables which will be the focus of this paper are the $s$-wave
pairing and charge density wave (CDW) correlation functions,
\begin{equation}
  \begin{aligned}
    \swave(\vec{i},\vec{j}) &=
    \langle \Delta^{}_{\vec{i}+\vec{j}} \Delta^{\dagger}_{\vec{i}} \rangle,
    \\
    \cdw(\vec{i},\vec{j}) &=
    \langle n_{\vec{i}+\vec{j}} n_{\vec{i}} \rangle 
    - \langle n_{\vec{i}+\vec{j}} \rangle \langle n_{\vec{i}} \rangle,
  \end{aligned}
\end{equation}
where $\Delta^{\dagger}_{\vec{i}} = c^{\dagger}_{\vec{i}\uparrow}
c^{\dagger}_{\vec{i}\downarrow}$ creates a \emph{pair} of fermions on
site $\vec i$ and $n_{\vec i} = n_{\vec{i}\uparrow} +
n_{\vec{i}\downarrow}$ counts the fermions on site $\vec i$.  Notice
that these depend on $\vec{i}$ and not just on the separation
$\vec{j}$.  We also define the associated structure factors
\begin{equation}
  \label{eq:SF}
  \begin{aligned}
    \strfact &= \sum_{\vec i \vec j} \swave(\vec i, \vec j),
    \\
    S_\text{cdw} &= \sum_{\vec i \vec j} (-1)^{\vec j}\,
    \cdw(\vec i, \vec j).
  \end{aligned}
\end{equation}
In addition, we study the local quantities
\begin{equation}
  \begin{aligned}
    \density(\vec i) &= \langle n_{\vec i \uparrow} \rangle +
    \langle n_{\vec i \downarrow} \rangle,
    \\
    \kine(\vec i) &= \tfrac12 \sum_{\vec j}^{\langle \vec i\vec j \rangle} 
    ( c^{\dagger}_{\vec{i}\sigma} c^{}_{\vec{j}\sigma} +
    c^{\dagger}_{\vec{j}\sigma} c^{}_{\vec{i}\sigma} ),
    \\
    \docc(\vec i) &=
    \langle n_{\vec i \uparrow} n_{\vec i \downarrow}\rangle -
    \langle n_{\vec i \uparrow} \rangle\langle n_{\vec i \downarrow}\rangle;
  \end{aligned}
\end{equation}
the density $\density(\vec i)$ has already been used; $\kine(\vec i)$
is the kinetic energy associated with the bonds of site $\vec i$; and
$\docc(\vec i)$ is the double occupancy with the trivial density
dependence subtracted.

Before proceeding with the confined case, it is useful to review the
properties of the translationally invariant case, $\vtrap=0$.  In two
dimensions, it is known that the half-filled attractive Hubbard
Hamiltonian has combined long-range CDW and $s$-wave pairing order in
its ground state, and is unordered at nonzero
temperature.\cite{negurev} When doped, the symmetry between charge and
pairing is broken, and a Kosterlitz-Thouless (KT) transition to a
quasi-long-range ordered superfluid phase occurs at finite
temperature.  Many numerical and analytical studies have been
performed.\cite{negutc} The transition temperature $\tcrit$ rises
rapidly as $\mu$ is made non-zero and reaches a maximum value of
$\tcrit \simeq \nicefrac{1}{10}$ for a wide range of fillings $0.5 <
\density < 0.9$.\cite{paiva04} The effect of inhomogeneities in the
interaction strength has also been explored.  \cite{aryanpour07}

Consideration of the ``asymmetric'' particle-hole (PH) transformation
helps clarify these assertions.  On a bipartite lattice, when the
down-spin operations in the AHM, \eqref{eq:hamiltonian} are mapped
with
\begin{equation}
  \label{eq:PH}
  c^{}_{\vec{i}\downarrow} \leftrightarrow
  (-1)^{\vec i} c^{\dagger}_{\vec{i}\downarrow},
\end{equation}
the kinetic energy is unchanged, but the interaction term changes
sign, so that the AHM maps onto the RHM (the phase factor $(-1)^{\vec
  i}$ is understood to take the values $\pm1$ on alternating
sublattices).  This PH symmetry provides a simple argument that the
two-dimensional half-filled AHM can have long-range order (LRO) only
at $T=0$, like the RHM.

QMC simulations have shown that the ground state of the half-filled,
two-dimensional uniform RHM is magnetically ordered.\cite{hirsch85} PH
symmetry then implies that CDW and pair order occur simultaneously in
the $T=0$ half-filled AHM.  To see this, note that the $z$ component
of spin $n_{\vec{i}\uparrow} - n_{\vec{i}\downarrow}$ in the RHM maps
onto the charge $n_{\vec{i}\uparrow} + n_{\vec{i}\downarrow}$ in the
AHM, so that magnetic LRO in the $z$ direction of the RHM corresponds
to CDW order of the AHM.  Similarly, magnetic order in the $xy$ plane
maps onto $s$-wave pairing order.  The degeneracy of the $z$ and $xy$
magnetic order in the repulsive model implies that CDW and pair order
occur simultaneously in the half-filled attractive case.

A final consequence of PH symmetry is the explanation of the
occurrence of pairing order (and the absence of CDW order) at finite
temperature in the doped AHM.  When doped, $\mu$ is non-zero.  Under
the PH transformation \eqref{eq:PH}, the chemical potential term $\mu
(n_{\vec{i}\uparrow} + n_{\vec{i}\downarrow})$ in the AHM becomes a
Zeeman field $B (n_{\vec{i}\uparrow} - n_{\vec{i}\downarrow})$ in the
RHM ($B$ has the same numerical value as $\mu$).  Because the order in
the RHM is antiferromagnetic, a uniform field in the $z$ direction
makes it energetically favorable for spins to lie in the $xy$ plane,
since then they can tilt out of the plane and pick up field energy
without costing as much exchange energy.  This lowering of symmetry
from three to two components makes possible a finite temperature
Kosterlitz-Thouless transition in two dimensions.  The $xy$ magnetic
order which exists in the RHM then maps to $s$-wave pair order in the
AHM.

\section*{Correlations and the Local Density Approximation}
%

\begin{figure}[tbp]
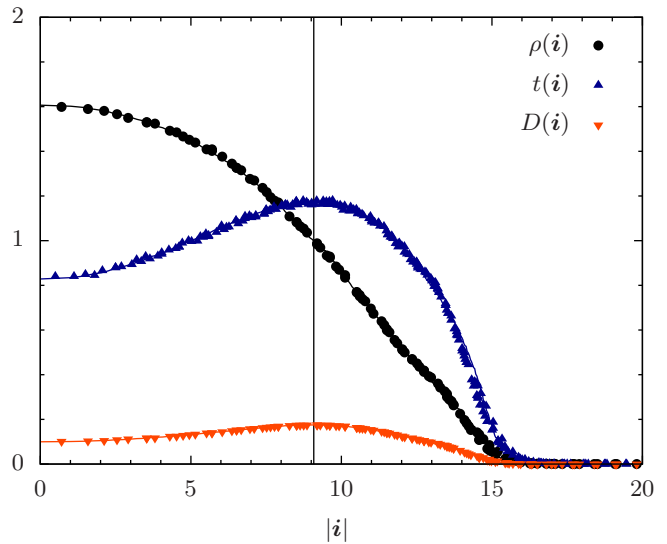

  \centering
  \include{30x30_dens}

  \caption{(Color online) Fermion density \density, kinetic energy
    \kine and double occupancy \docc as a function of distance \dist
    from the trap center, with $\npart=564$ particles, $U=6$, $\vtrap
    = 0.0097$ and $\mu=0.8$ at $\beta=9$.  The solid line shows the
    LDA result; in this case, near perfect agreement is found between
    the true trapped system and the LDA.  Note that, in contrast to
    the RHM, there is no plateau at half-filling (marked by the
    vertical line), which corresponds to the absence of a Mott gap in
    the homogeneous model.  Error bars are smaller than the symbols.}
  \label{dens}
\end{figure}

We begin by showing the density profile in \figref{dens}, along with
the local kinetic energy and the double occupancy.  Results are given
both from the LDA and from a trapped $30 \times 30$ system with
$\npart=564.1\pm0.4$ particles.  These two approaches yield results in very
good agreement for $\density(\vec i)$, $\kine(\vec i)$ and $\docc(\vec
i)$.  An important point is the absence of a density plateau at
$\density=1$, in accordance with results in the LDA, and also with a
particle-hole symmetry argument which identifies the compressibility
$\kappa=d\density/d\mu$ of the AHM with the uniform magnetic
susceptibility $\chi=dM/dB$ of the RHM, which is known to be nonzero
at zero external field; thus, as noted above, the AHM has finite
compressibility at integer filling, and there is no Mott plateau at
half-filling.
 
This true one-dimensionality of the $\density(\vec i)=1$ ring makes
the formation of long range CDW order in the AHM much less robust than
the antiferromagnetic order which can occur on the
quasi-two-dimensional Mott annulus of integer filling that occurs in
the RHM.

A related difference to the RHM is seen in the kinetic energy, which
shows a maximum at half-filling in \figref{dens}, where in the RHM the
Mott phase would lead to localization and a minimum of the kinetic
energy.  This behavior is best understood by applying the asymmetric
PH transformation.  The corresponding RHM is uniformly half-filled and
subject to a perpendicular Zeeman field $B_{\vec i}$ that varies
radially and goes through zero at those sites $\vec i$ that had
$\rho(\vec i)=1$ in the original, attractive model.  Away from the
$B=0$ region, the system gets increasingly spin-polarized and thereby
makes Pauli exclusion more effective in hindering fermion mobility.
One therefore expects a maximum in both $|\kine(\vec i)|$ and
$|\docc(\vec i)|$ when the site $\vec i$ belongs to the $B_{\vec i}=0$
region.  As these quantities are unchanged by the PH transformation,
this last statement translates \emph{verbatim} to the AHM.


\begin{figure}[tbp]
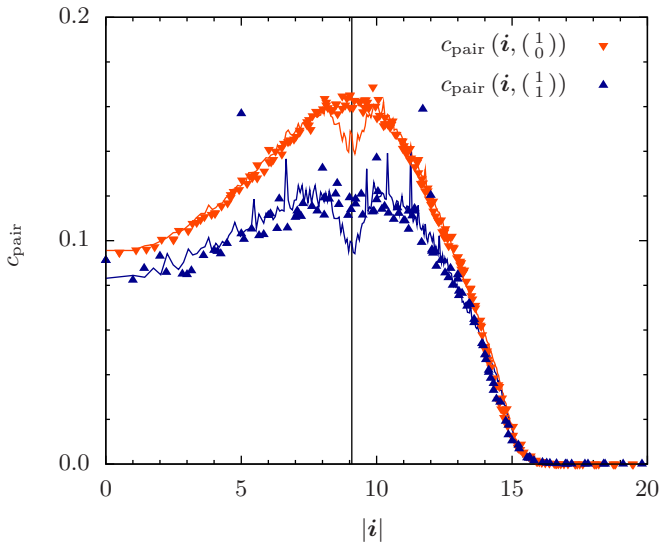

  \centering
  \include{30x30_swave}

  \caption{(Color online) Near-neighbor [\swnearn{}] and next-near
    neighbor [\swnextn{}] $s$-wave pair correlations as a function of
    distance \dist from the trap center, with $\npart=564$ particles,
    $U=6$, $\vtrap = 0.0097$ and $\mu=0.8$ at $\beta=9$.  The solid
    lines show the LDA result; in this case, a striking failure of the
    LDA is seen around half-filling (marked by the vertical line).
    Due to CDW/$s$-wave degeneracy at half-filling in the homogeneous
    model, the LDA predicts superfluidity only sufficiently far from
    that point, while in the trapped system the superfluid phase may
    penetrate the half-filled ring.  Error bars have been suppressed
    to avoid clutter, but the spread of the data points gives an
    indication of the uncertainties, which result from large
    statistical fluctuations observed at low $T$.  (See text.)}
\label{swave}
\end{figure}

\begin{figure}[tbp]
  \centering
  \include{30x30_cdw}

  \caption{(Color online) Near-neighbor [\swnearn{}] and next-near
    neighbor [\swnextn, with the sign inverted for clarity] CDW
    correlations as a function of distance \dist from the trap center,
    with $N=564$ particles, $U=6$, $\vtrap = 0.0097$ and $\mu=0.8$ at
    $\beta=9$.  The solid lines show the LDA result.  As with the pair
    correlation, the LDA fails around half-filling (marked by the
    vertical line), where it predicts enhanced CDW correlations.
    Error bars are smaller than the symbols.}
\label{cdw}
\end{figure}

Next, \figref{swave} shows the near-neighbor $\swnearn$ and next-near
neighbor $\swnextn$ $s$-wave pairing correlators both in the LDA and
the $30 \times 30$ system.  In the LDA, both functions dip at
$r=\sqrt{\mu/\vtrap}$, where the local density $\density(\vec i)=1$,
as do the corresponding farthest-neighbor correlators.  \figref{cdw}
shows the density correlators $\cdwnearn$ and $|\cdwnextn|$ which in the
LDA peak as the system crosses through commensurate filling.

The dip (peak) in the pairing (CDW) correlation functions observed in
the LDA may be understood from the CDW-pairing degeneracy that exists
precisely at half-filling, and the corresponding suppression of the
finite-$T$ pairing order that exists away from half-filling.

While the LDA compares favorably with the \emph{ab initio} calculation
across most of the lattice, the dip (peak) in the $s$-wave pairing
(CDW) when the $\density(\vec i)=1$ ring is crossed is conspicuously
absent in the true trapped system.

It is useful to compare this behavior with the RHM, where the physics
of commensurate filling ($\density=1$) can be inferred correctly, for
the most part, from the LDA because of the presence of an annulus of
finite thickness which ``protects'' the Mott region.  By contrast, in
the AHM, there is no such protection; the half-filled ring is truly
one-dimensional, and the physics of commensurate filling is
essentially absent in the trap.

Once again, a deeper understanding can be reached by applying the PH
transformation: the $B=0$ region of the corresponding RHM divides, and
is coupled to, regions where the spins are tilted out of the $xy$
plane in opposite directions.  Spins in the $B=0$ region can then
lower the system's energy by aligning with neighboring spins on the
$xy$ plane and therefore breaking the local $SU(2)$ symmetry
characteristic of the Hubbard model in the absence of an external
field.  In the original language of the AHM this implies that, at
half-filling, one should expect a reduction of the CDW correlation, an
increase in the pairing correlations and a breaking of the CDW-pairing
degeneracy.

For optical-lattice experiments that aim to emulate the Hubbard model,
our findings indicate that the physics of the half-filled AHM will be
inaccessible in any experimental setup that leads to a confining
potential $(\mu - \vtrap\dist^2)$ as in \eqref{eq:hamiltonian}.  It
has recently been suggested\cite{ho09} to simulate the AHM on an
optical lattice, and then utilize the PH transformation \eqref{eq:PH}
to draw conclusions about the repulsive case.  However, this proposal
appears to be challenging, since the effective absence of the
half-filled case in the confined AHM means that the physics of the RHM
at zero Zeeman field will be inaccessible.

Both problems outlined above could be solved by the putative
``off-diagonal'' confinement (ODC) \cite{odc}, if realized, since
particle-hole symmetry indicates that the inhomogeneous lattice can be
made uniformly half-filled under ODC.  On the other hand, in the
conventional ``diagonal confinement'', observing a finite-temperature
transition to a superconducting phase in confined systems becomes much
more likely when the CDW region only occupies such a limited spatial
region, as discussed in detail in the following section.

\section*{Finite-Size Extrapolation}

We now turn to the interesting issue of the finite-$T$ phase
transition in the confined AHM.  In general we may ask, when a
trapping potential is added to a model that undergoes, e.g.~a KT
transition in the homogeneous case, how is the nature of that phase
transition altered by the trap?  In the case of the classical
$xy$-model, it has been shown\cite{crecchi11} that the KT transition
of the homogeneous model is preserved in many respects in the trapped
case.  In this section, we present a finite-size extrapolation (FSE)
to address the same question for the AHM.

True phase transitions of course can occur only in the thermodynamic
limit of infinite system size.  In a translationally invariant system,
the correct way to perform this limit is familiar and almost trivial:
the global density $\density = \npart/L^2$ is kept constant.  In the
presence of a trap, the correct generalization is to keep the the
``characteristic density'' $\rhoc := \npart / \Lc^2$ constant.\cite{rhoc}
Here $\Lc = \sqrt{t/\vtrap}$ is the natural length scale in the
problem, formed by combining the kinetic energy $t$ and the trap
curvature $\vtrap$.  In the finite-size analysis described below, we
have used systems of constant $\rhoc$.

Of course, since only finite system sizes are accessible numerically,
one must invoke a procedure to infer behavior in the thermodynamic
limit from finite lattices.  Here, we follow the finite-size scaling
(FSS) procedure discussed in \citeref{paiva04}.  The generalization of
FSS in the presence of the trapping potential has been called
\emph{trap-size scaling} (TSS)\cite{campostrini09}; we note that TSS
may be expressed as keeping $\rhoc$ constant and proceeding with the
FSS analysis as usual.

\begin{figure}[tbp]
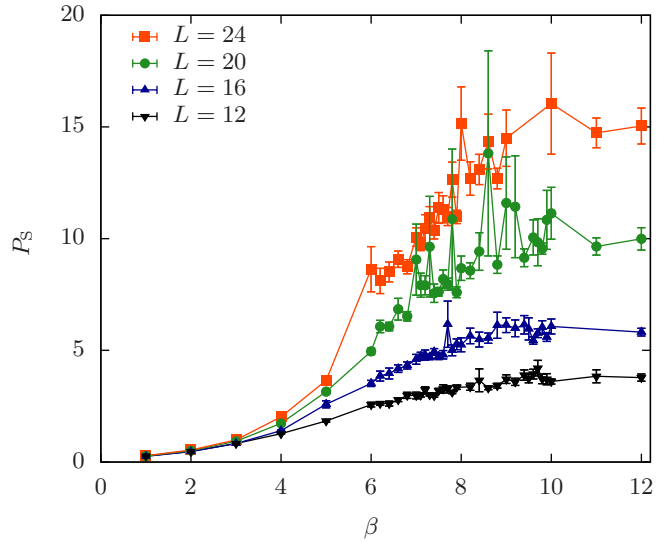

  \centering
  \include{fss_pstr}
  \caption{(Color online) The $s$-wave structure factor \strfact for
    systems of different linear size $\lattsz$ with $U=6$ as a
    function of inverse temperature $\beta$.  At low temperatures
    ($\beta \gtrsim 7$), the curves flatten off to values determined
    by the system size, indicating a divergent correlation length.}
  \label{fss_plain}
\end{figure}

In this approach, the pair structure factor \strfact of \eqref{eq:SF}
is obtained for different lattice sizes and temperatures (plotted in
\figref{fss_plain}).  We see that at high temperatures ($\beta
\lesssim 4$), when the correlation length is short, \strfact is
independent of the system size.  At lower temperatures ($\beta \simeq
4$), the curves begin to separate as the correlation length becomes
large compared to the system sizes.  Finally, for very low
temperatures \strfact approaches a constant depending on the system
size.  This behavior is expected for KT and second order transitions
where observables stop evolving with temperature when the correlation
length $\xi$ exceeds $L$.
 
Unfortunately, the full FSS procedure of \citeref{paiva04} is defeated
in the case of the confined AHM because of excessive statistical
fluctuations in the DQMC estimator for \swave.  
These large fluctuations are manifest in, for example, the
very large error bars at in two of  the near-neighbor
pair correlators of Fig.~\ref{swave}.
However, a comparison
of results on lattice sizes $12,\,14,\ldots,24$ at temperatures $1 \le
\beta \le 16$ provides evidence that the KT transition is indeed
preserved in the trap.
The approach hinges on the critical scaling of the structure factor
near the phase transition; the expected KT form is
\begin{equation}
  \label{eq:SFscaling}
  \strfact \sim L^{2-\eta(T)}
\end{equation}
with $\eta(\tcrit)=\nicefrac14$.  The critical exponent $\eta$ for a
KT transition in a homogeneous system is known to vary with
temperature between $\eta(T=0)=0$ and
$\eta(\tcrit)=\nicefrac14$.\cite{etaKT} In the trapped system, this
issue is complicated by the varying filling.  Arguing within the LDA,
since the filling varies in the lattice, so does \tcrit; but $\eta$
must be a function of $T/\tcrit$ rather than $T$ itself, therefore
$\eta$ should vary across the system along with the filling.  We must
therefore ask whether an effective exponent \effeta exists such that
\eqref{eq:SFscaling} holds for the system as a whole.

\begin{figure}[tbp]
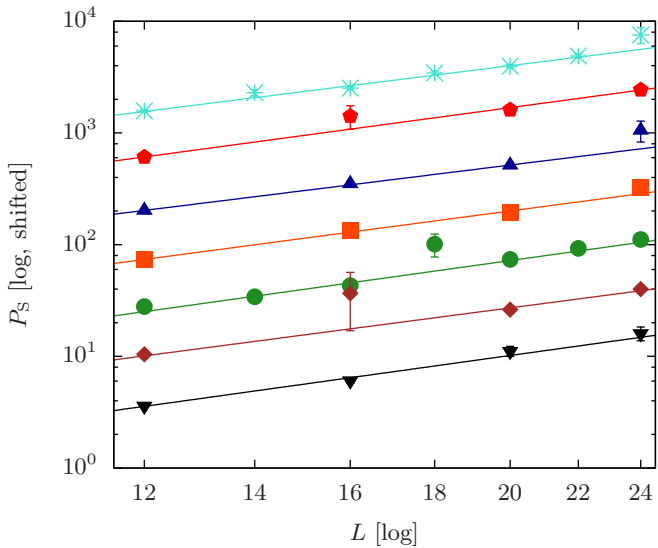

  \centering
  \include{etafit_struct_off}
  \caption{(Color online) The $s$-wave structure factor \strfact as a
    function of linear lattice size $\lattsz$ for $U=6$ at various low
    temperatures (top to bottom: $\beta = \nicefrac1T = 16, 15,
    \ldots, 10$) on a doubly logarithmic scale.  A vertical shift has
    been applied to separate the curves from each other.  A linear
    dependence on the doubly logarithmic scale (i.e., the structure
    factor varies as a power of the system size) points to a divergent
    correlation length.}
  \label{fss_loglog}
\end{figure}

To investigate the behavior of $\eta$ in the trapped system, we plot
$\log\strfact$ as a function of $\log\lattsz$ for several low
temperatures, $\beta \ge 10$, in \figref{fss_loglog}.  A straight
line with slope $2-\effeta$ in the doubly logarithmic plot indicates
that the scaling relation \eqref{eq:SFscaling} holds for the whole
system with an effective exponent \effeta.  We have used non-linear
least-squares fits to estimate \effeta at temperatures $\beta \ge
6.6$, using the data shown in \figref{fss_plain}.  Within the
statistical uncertainties, we find good agreement with
\eqref{eq:SFscaling}; some examples of the fits are shown in
\figref{fss_loglog}.

As a function of temperature, the behavior of \effeta is consistent
with the behavior expected from the KT transition in the homogeneous
case, dropping off to zero as the temperature is decreased.  Between
$\beta=8 \text{--} 10$, we find large fluctuations of \effeta, which
may be interpreted as a signature of the phase transition.  Thus, we
may estimate the critical temperature of the system as $\tcrit
\simeq 0.1\text{--}0.125$.

Our results for \effeta are consistent with recent work on the
classical $xy$-model\cite{crecchi11}, where it was found that \effeta
for the trapped system at a certain temperature takes the same value
as $\eta$ in the homogeneous system at the same temperature.


\section*{Conclusions}

In this paper, the attractive Hubbard model (AHM) in a harmonic
confining potential was studied, especially with regards to
superfluidity at low temperatures.  Results from the local-density
approximation (LDA) were compared to calculations within the true
trapped system.

While the LDA is valid for local observables in most cases, we find
qualitatively wrong predictions for the $s$-wave pairing and CDW
correlation functions around the ring of half-filling, where the LDA
predicts a dip and a peak, respectively, which are absent when the
trapped system is treated ab initio.  This is linked to the
relationship between the pairing and CDW correlations at half-filling;
and to the density profile in the trap, where the Mott plateau
exhibited by the repulsive Hubbard model is absent.  Consequently, the
physics of the half-filled case is not represented anywhere in the
trap.  On the other hand, that the CDW correlations are suppressed
will make it easier to observe a transition to a superfluid phase at
finite temperature.

A finite-size extrapolation, where systems of different sizes but at
the same characteristic density \rhoc are compared, provides evidence
that the Kosterlitz-Thouless transition to a superfluid phase of the
homogeneous AHM persists in the trap, with a critical temperature of
$\tcrit \simeq 0.1\text{--}0.125$.

\begin{acknowledgments}
  We are grateful to E.~Vicari, T.~Paiva, F.~Assaad and L.~Hopkins for
  instructive conversations.  E.A.\ gratefully acknowledges support
  from the Austrian Marshall Plan Foundation.  Supported under a
  CNRS-UC Davis EPOCAL LIA joint research grant, ARO Award
  W911NF0710576 with funds from the DARPA OLE Program, and by the
  Department of Energy under DE-FG52-09NA2946.
\end{acknowledgments}

\end{document}

%% file: definitions.tex
\newcommand{\density}{\ensuremath{\rho}\xspace}
\newcommand{\rhoc}{\ensuremath{\tilde\rho}\xspace}
\newcommand{\Lc}{\ensuremath{\ell}\xspace}
\newcommand{\swave}{\ensuremath{c_\text{pair}}\xspace}
\newcommand{\swnearn}{\ensuremath{\swave\left(\vec i, \left(
    \begin{smallmatrix} 1 \\ 0 \end{smallmatrix}
    \right)\right)}\xspace}
\newcommand{\swnextn}{\ensuremath{\swave\left(\vec i, \left(
    \begin{smallmatrix} 1 \\ 1 \end{smallmatrix}
    \right)\right)}\xspace}
\newcommand{\cdw}{\ensuremath{c_\text{charge}}\xspace}
\newcommand{\cdwnearn}{\ensuremath{\cdw\left(\vec i, \left(
    \begin{smallmatrix} 1 \\ 0 \end{smallmatrix}
    \right)\right)}\xspace}
\newcommand{\cdwnextn}{\ensuremath{\cdw\left(\vec i, \left(
    \begin{smallmatrix} 1 \\ 1 \end{smallmatrix}
    \right)\right)}\xspace}
\newcommand{\invtemp}{\ensuremath{\beta}\xspace}

\newcommand{\tcrit}{\ensuremath{T_\text{c}}\xspace}

\newcommand{\vtrap}{\ensuremath{V_\text{trap}}\xspace}
\newcommand{\dist}{\ensuremath{|\vec i|}\xspace}

\newcommand{\strfact}{\ensuremath{P_\text{S}}\xspace}
\newcommand{\lattsz}{\ensuremath{L}\xspace}
\newcommand{\effeta}{\ensuremath{\tilde\eta}\xspace}
\newcommand{\docc}{\ensuremath{D}\xspace}

\newcommand{\kine}{\ensuremath{t}\xspace}
\newcommand{\npart}{N}

\newcommand{\figref}[1]{Fig.~\ref{#1}}
\newcommand{\citeref}[1]{[\onlinecite{#1}]}
\renewcommand\eqref[1]{Eq.~\ref{#1}}

\DeclareMathOperator{\trace}{Tr}
\renewcommand{\vec}[1]{\boldsymbol{#1}}



%% file: 30x30_dens.tex
\begingroup
  \makeatletter
  \providecommand\color[2][]{%
    \GenericError{(gnuplot) \space\space\space\@spaces}{%
      Package color not loaded in conjunction with
      terminal option `colourtext'%
    }{See the gnuplot documentation for explanation.%
    }{Either use 'blacktext' in gnuplot or load the package
      color.sty in LaTeX.}%
    \renewcommand\color[2][]{}%
  }%
  \providecommand\includegraphics[2][]{%
    \GenericError{(gnuplot) \space\space\space\@spaces}{%
      Package graphicx or graphics not loaded%
    }{See the gnuplot documentation for explanation.%
    }{The gnuplot epslatex terminal needs graphicx.sty or graphics.sty.}%
    \renewcommand\includegraphics[2][]{}%
  }%
  \providecommand\rotatebox[2]{#2}%
  \@ifundefined{ifGPcolor}{%
    \newif\ifGPcolor
    \GPcolortrue
  }{}%
  \@ifundefined{ifGPblacktext}{%
    \newif\ifGPblacktext
    \GPblacktexttrue
  }{}%
  \let\gplgaddtomacro\g@addto@macro
  \gdef\gplbacktext{}%
  \gdef\gplfronttext{}%
  \makeatother
  \ifGPblacktext
    \def\colorrgb#1{}%
    \def\colorgray#1{}%
  \else
    \ifGPcolor
      \def\colorrgb#1{\color[rgb]{#1}}%
      \def\colorgray#1{\color[gray]{#1}}%
      \expandafter\def\csname LTw\endcsname{\color{white}}%
      \expandafter\def\csname LTb\endcsname{\color{black}}%
      \expandafter\def\csname LTa\endcsname{\color{black}}%
      \expandafter\def\csname LT0\endcsname{\color[rgb]{1,0,0}}%
      \expandafter\def\csname LT1\endcsname{\color[rgb]{0,1,0}}%
      \expandafter\def\csname LT2\endcsname{\color[rgb]{0,0,1}}%
      \expandafter\def\csname LT3\endcsname{\color[rgb]{1,0,1}}%
      \expandafter\def\csname LT4\endcsname{\color[rgb]{0,1,1}}%
      \expandafter\def\csname LT5\endcsname{\color[rgb]{1,1,0}}%
      \expandafter\def\csname LT6\endcsname{\color[rgb]{0,0,0}}%
      \expandafter\def\csname LT7\endcsname{\color[rgb]{1,0.3,0}}%
      \expandafter\def\csname LT8\endcsname{\color[rgb]{0.5,0.5,0.5}}%
    \else
      \def\colorrgb#1{\color{black}}%
      \def\colorgray#1{\color[gray]{#1}}%
      \expandafter\def\csname LTw\endcsname{\color{white}}%
      \expandafter\def\csname LTb\endcsname{\color{black}}%
      \expandafter\def\csname LTa\endcsname{\color{black}}%
      \expandafter\def\csname LT0\endcsname{\color{black}}%
      \expandafter\def\csname LT1\endcsname{\color{black}}%
      \expandafter\def\csname LT2\endcsname{\color{black}}%
      \expandafter\def\csname LT3\endcsname{\color{black}}%
      \expandafter\def\csname LT4\endcsname{\color{black}}%
      \expandafter\def\csname LT5\endcsname{\color{black}}%
      \expandafter\def\csname LT6\endcsname{\color{black}}%
      \expandafter\def\csname LT7\endcsname{\color{black}}%
      \expandafter\def\csname LT8\endcsname{\color{black}}%
    \fi
  \fi
  \setlength{\unitlength}{0.0500bp}%
  \begin{picture}(5102.00,4250.00)%
    \gplgaddtomacro\gplbacktext{%
      \csname LTb\endcsname%
      \put(144,640){\makebox(0,0)[r]{\strut{}0}}%
      \put(144,2325){\makebox(0,0)[r]{\strut{}1}}%
      \put(144,4010){\makebox(0,0)[r]{\strut{}2}}%
      \put(264,440){\makebox(0,0){\strut{}$0$}}%
      \put(1399,440){\makebox(0,0){\strut{}$5$}}%
      \put(2533,440){\makebox(0,0){\strut{}$10$}}%
      \put(3668,440){\makebox(0,0){\strut{}$15$}}%
      \put(4802,440){\makebox(0,0){\strut{}$20$}}%
      \put(2533,140){\makebox(0,0){\strut{}\dist}}%
    }%
    \gplgaddtomacro\gplfronttext{%
      \csname LTb\endcsname%
      \put(4259,3797){\makebox(0,0)[r]{\strut{}$\density(\vec i)$}}%
      \csname LTb\endcsname%
      \put(4259,3497){\makebox(0,0)[r]{\strut{}$\kine(\vec i)$}}%
      \csname LTb\endcsname%
      \put(4259,3197){\makebox(0,0)[r]{\strut{}$\docc(\vec i)$}}%
    }%
    \gplbacktext
    \put(0,0){\includegraphics{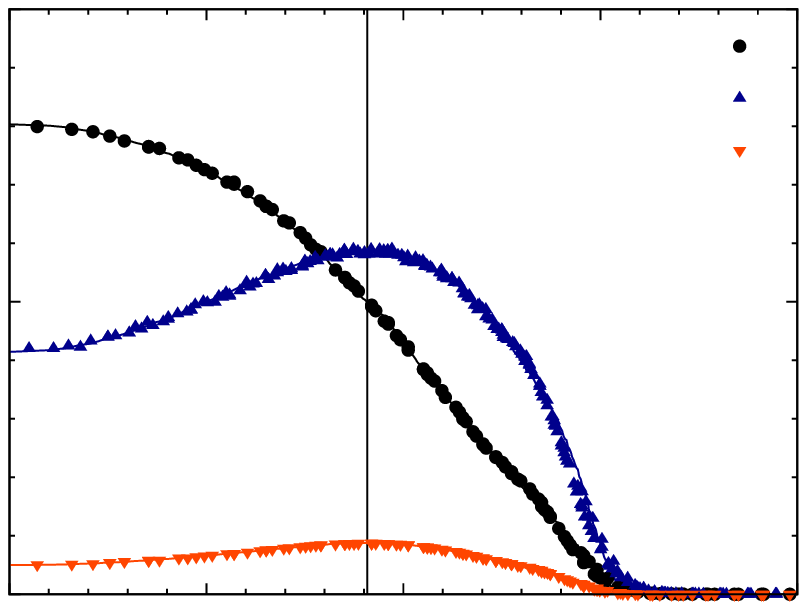}}%
    \gplfronttext
  \end{picture}%
\endgroup

%% file: 30x30_swave.tex
\begingroup
  \makeatletter
  \providecommand\color[2][]{%
    \GenericError{(gnuplot) \space\space\space\@spaces}{%
      Package color not loaded in conjunction with
      terminal option `colourtext'%
    }{See the gnuplot documentation for explanation.%
    }{Either use 'blacktext' in gnuplot or load the package
      color.sty in LaTeX.}%
    \renewcommand\color[2][]{}%
  }%
  \providecommand\includegraphics[2][]{%
    \GenericError{(gnuplot) \space\space\space\@spaces}{%
      Package graphicx or graphics not loaded%
    }{See the gnuplot documentation for explanation.%
    }{The gnuplot epslatex terminal needs graphicx.sty or graphics.sty.}%
    \renewcommand\includegraphics[2][]{}%
  }%
  \providecommand\rotatebox[2]{#2}%
  \@ifundefined{ifGPcolor}{%
    \newif\ifGPcolor
    \GPcolortrue
  }{}%
  \@ifundefined{ifGPblacktext}{%
    \newif\ifGPblacktext
    \GPblacktexttrue
  }{}%
  \let\gplgaddtomacro\g@addto@macro
  \gdef\gplbacktext{}%
  \gdef\gplfronttext{}%
  \makeatother
  \ifGPblacktext
    \def\colorrgb#1{}%
    \def\colorgray#1{}%
  \else
    \ifGPcolor
      \def\colorrgb#1{\color[rgb]{#1}}%
      \def\colorgray#1{\color[gray]{#1}}%
      \expandafter\def\csname LTw\endcsname{\color{white}}%
      \expandafter\def\csname LTb\endcsname{\color{black}}%
      \expandafter\def\csname LTa\endcsname{\color{black}}%
      \expandafter\def\csname LT0\endcsname{\color[rgb]{1,0,0}}%
      \expandafter\def\csname LT1\endcsname{\color[rgb]{0,1,0}}%
      \expandafter\def\csname LT2\endcsname{\color[rgb]{0,0,1}}%
      \expandafter\def\csname LT3\endcsname{\color[rgb]{1,0,1}}%
      \expandafter\def\csname LT4\endcsname{\color[rgb]{0,1,1}}%
      \expandafter\def\csname LT5\endcsname{\color[rgb]{1,1,0}}%
      \expandafter\def\csname LT6\endcsname{\color[rgb]{0,0,0}}%
      \expandafter\def\csname LT7\endcsname{\color[rgb]{1,0.3,0}}%
      \expandafter\def\csname LT8\endcsname{\color[rgb]{0.5,0.5,0.5}}%
    \else
      \def\colorrgb#1{\color{black}}%
      \def\colorgray#1{\color[gray]{#1}}%
      \expandafter\def\csname LTw\endcsname{\color{white}}%
      \expandafter\def\csname LTb\endcsname{\color{black}}%
      \expandafter\def\csname LTa\endcsname{\color{black}}%
      \expandafter\def\csname LT0\endcsname{\color{black}}%
      \expandafter\def\csname LT1\endcsname{\color{black}}%
      \expandafter\def\csname LT2\endcsname{\color{black}}%
      \expandafter\def\csname LT3\endcsname{\color{black}}%
      \expandafter\def\csname LT4\endcsname{\color{black}}%
      \expandafter\def\csname LT5\endcsname{\color{black}}%
      \expandafter\def\csname LT6\endcsname{\color{black}}%
      \expandafter\def\csname LT7\endcsname{\color{black}}%
      \expandafter\def\csname LT8\endcsname{\color{black}}%
    \fi
  \fi
  \setlength{\unitlength}{0.0500bp}%
  \begin{picture}(5102.00,4250.00)%
    \gplgaddtomacro\gplbacktext{%
      \csname LTb\endcsname%
      \put(600,640){\makebox(0,0)[r]{\strut{}0.0}}%
      \put(600,2325){\makebox(0,0)[r]{\strut{}0.1}}%
      \put(600,4010){\makebox(0,0)[r]{\strut{}0.2}}%
      \put(720,440){\makebox(0,0){\strut{}$0$}}%
      \put(1741,440){\makebox(0,0){\strut{}$5$}}%
      \put(2761,440){\makebox(0,0){\strut{}$10$}}%
      \put(3782,440){\makebox(0,0){\strut{}$15$}}%
      \put(4802,440){\makebox(0,0){\strut{}$20$}}%
      \put(20,2325){\rotatebox{90}{\makebox(0,0){\strut{}\swave}}}%
      \put(2761,140){\makebox(0,0){\strut{}\dist}}%
    }%
    \gplgaddtomacro\gplfronttext{%
      \csname LTb\endcsname%
      \put(4259,3797){\makebox(0,0)[r]{\strut{}\swnearn}}%
      \csname LTb\endcsname%
      \put(4259,3497){\makebox(0,0)[r]{\strut{}\swnextn}}%
    }%
    \gplbacktext
    \put(0,0){\includegraphics{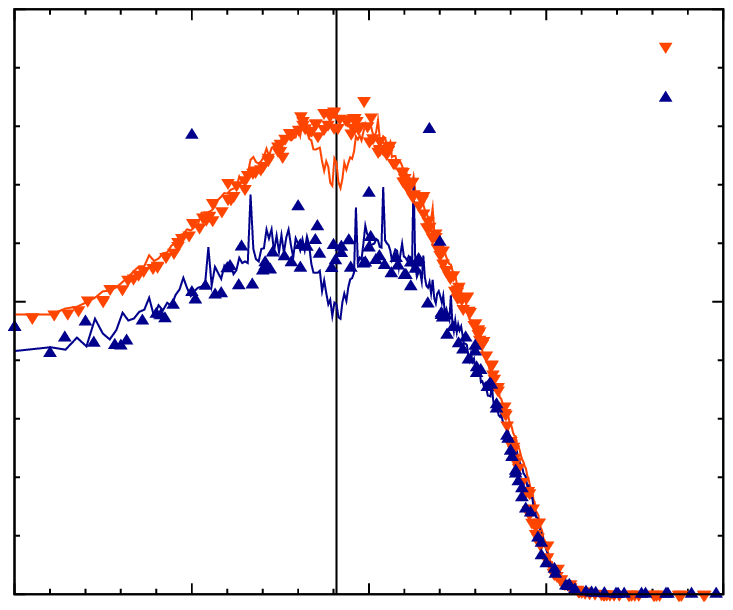}}%
    \gplfronttext
  \end{picture}%
\endgroup

%% file: 30x30_cdw.tex
\begingroup
  \makeatletter
  \providecommand\color[2][]{%
    \GenericError{(gnuplot) \space\space\space\@spaces}{%
      Package color not loaded in conjunction with
      terminal option `colourtext'%
    }{See the gnuplot documentation for explanation.%
    }{Either use 'blacktext' in gnuplot or load the package
      color.sty in LaTeX.}%
    \renewcommand\color[2][]{}%
  }%
  \providecommand\includegraphics[2][]{%
    \GenericError{(gnuplot) \space\space\space\@spaces}{%
      Package graphicx or graphics not loaded%
    }{See the gnuplot documentation for explanation.%
    }{The gnuplot epslatex terminal needs graphicx.sty or graphics.sty.}%
    \renewcommand\includegraphics[2][]{}%
  }%
  \providecommand\rotatebox[2]{#2}%
  \@ifundefined{ifGPcolor}{%
    \newif\ifGPcolor
    \GPcolortrue
  }{}%
  \@ifundefined{ifGPblacktext}{%
    \newif\ifGPblacktext
    \GPblacktexttrue
  }{}%
  \let\gplgaddtomacro\g@addto@macro
  \gdef\gplbacktext{}%
  \gdef\gplfronttext{}%
  \makeatother
  \ifGPblacktext
    \def\colorrgb#1{}%
    \def\colorgray#1{}%
  \else
    \ifGPcolor
      \def\colorrgb#1{\color[rgb]{#1}}%
      \def\colorgray#1{\color[gray]{#1}}%
      \expandafter\def\csname LTw\endcsname{\color{white}}%
      \expandafter\def\csname LTb\endcsname{\color{black}}%
      \expandafter\def\csname LTa\endcsname{\color{black}}%
      \expandafter\def\csname LT0\endcsname{\color[rgb]{1,0,0}}%
      \expandafter\def\csname LT1\endcsname{\color[rgb]{0,1,0}}%
      \expandafter\def\csname LT2\endcsname{\color[rgb]{0,0,1}}%
      \expandafter\def\csname LT3\endcsname{\color[rgb]{1,0,1}}%
      \expandafter\def\csname LT4\endcsname{\color[rgb]{0,1,1}}%
      \expandafter\def\csname LT5\endcsname{\color[rgb]{1,1,0}}%
      \expandafter\def\csname LT6\endcsname{\color[rgb]{0,0,0}}%
      \expandafter\def\csname LT7\endcsname{\color[rgb]{1,0.3,0}}%
      \expandafter\def\csname LT8\endcsname{\color[rgb]{0.5,0.5,0.5}}%
    \else
      \def\colorrgb#1{\color{black}}%
      \def\colorgray#1{\color[gray]{#1}}%
      \expandafter\def\csname LTw\endcsname{\color{white}}%
      \expandafter\def\csname LTb\endcsname{\color{black}}%
      \expandafter\def\csname LTa\endcsname{\color{black}}%
      \expandafter\def\csname LT0\endcsname{\color{black}}%
      \expandafter\def\csname LT1\endcsname{\color{black}}%
      \expandafter\def\csname LT2\endcsname{\color{black}}%
      \expandafter\def\csname LT3\endcsname{\color{black}}%
      \expandafter\def\csname LT4\endcsname{\color{black}}%
      \expandafter\def\csname LT5\endcsname{\color{black}}%
      \expandafter\def\csname LT6\endcsname{\color{black}}%
      \expandafter\def\csname LT7\endcsname{\color{black}}%
      \expandafter\def\csname LT8\endcsname{\color{black}}%
    \fi
  \fi
  \setlength{\unitlength}{0.0500bp}%
  \begin{picture}(5102.00,4250.00)%
    \gplgaddtomacro\gplbacktext{%
      \csname LTb\endcsname%
      \put(600,1314){\makebox(0,0)[r]{\strut{}0.0}}%
      \put(600,2662){\makebox(0,0)[r]{\strut{}0.2}}%
      \put(600,4010){\makebox(0,0)[r]{\strut{}0.4}}%
      \put(720,440){\makebox(0,0){\strut{}$0$}}%
      \put(1741,440){\makebox(0,0){\strut{}$5$}}%
      \put(2761,440){\makebox(0,0){\strut{}$10$}}%
      \put(3782,440){\makebox(0,0){\strut{}$15$}}%
      \put(4802,440){\makebox(0,0){\strut{}$20$}}%
      \put(20,2325){\rotatebox{90}{\makebox(0,0){\strut{}\cdw}}}%
      \put(2761,140){\makebox(0,0){\strut{}\dist}}%
    }%
    \gplgaddtomacro\gplfronttext{%
      \csname LTb\endcsname%
      \put(4259,3797){\makebox(0,0)[r]{\strut{}\cdwnearn}}%
      \csname LTb\endcsname%
      \put(4259,3497){\makebox(0,0)[r]{\strut{}-\cdwnextn}}%
    }%
    \gplbacktext
    \put(0,0){\includegraphics{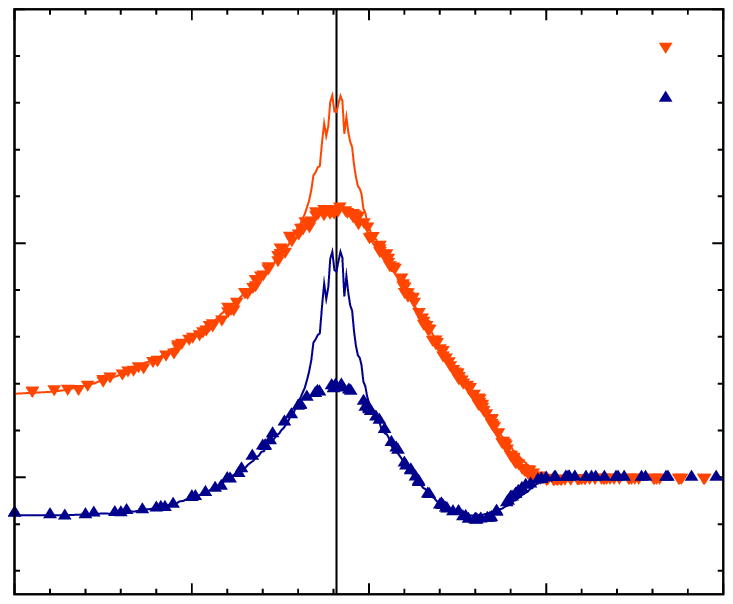}}%
    \gplfronttext
  \end{picture}%
\endgroup

%% file: fss_pstr.tex
\begingroup
  \makeatletter
  \providecommand\color[2][]{%
    \GenericError{(gnuplot) \space\space\space\@spaces}{%
      Package color not loaded in conjunction with
      terminal option `colourtext'%
    }{See the gnuplot documentation for explanation.%
    }{Either use 'blacktext' in gnuplot or load the package
      color.sty in LaTeX.}%
    \renewcommand\color[2][]{}%
  }%
  \providecommand\includegraphics[2][]{%
    \GenericError{(gnuplot) \space\space\space\@spaces}{%
      Package graphicx or graphics not loaded%
    }{See the gnuplot documentation for explanation.%
    }{The gnuplot epslatex terminal needs graphicx.sty or graphics.sty.}%
    \renewcommand\includegraphics[2][]{}%
  }%
  \providecommand\rotatebox[2]{#2}%
  \@ifundefined{ifGPcolor}{%
    \newif\ifGPcolor
    \GPcolortrue
  }{}%
  \@ifundefined{ifGPblacktext}{%
    \newif\ifGPblacktext
    \GPblacktexttrue
  }{}%
  \let\gplgaddtomacro\g@addto@macro
  \gdef\gplbacktext{}%
  \gdef\gplfronttext{}%
  \makeatother
  \ifGPblacktext
    \def\colorrgb#1{}%
    \def\colorgray#1{}%
  \else
    \ifGPcolor
      \def\colorrgb#1{\color[rgb]{#1}}%
      \def\colorgray#1{\color[gray]{#1}}%
      \expandafter\def\csname LTw\endcsname{\color{white}}%
      \expandafter\def\csname LTb\endcsname{\color{black}}%
      \expandafter\def\csname LTa\endcsname{\color{black}}%
      \expandafter\def\csname LT0\endcsname{\color[rgb]{1,0,0}}%
      \expandafter\def\csname LT1\endcsname{\color[rgb]{0,1,0}}%
      \expandafter\def\csname LT2\endcsname{\color[rgb]{0,0,1}}%
      \expandafter\def\csname LT3\endcsname{\color[rgb]{1,0,1}}%
      \expandafter\def\csname LT4\endcsname{\color[rgb]{0,1,1}}%
      \expandafter\def\csname LT5\endcsname{\color[rgb]{1,1,0}}%
      \expandafter\def\csname LT6\endcsname{\color[rgb]{0,0,0}}%
      \expandafter\def\csname LT7\endcsname{\color[rgb]{1,0.3,0}}%
      \expandafter\def\csname LT8\endcsname{\color[rgb]{0.5,0.5,0.5}}%
    \else
      \def\colorrgb#1{\color{black}}%
      \def\colorgray#1{\color[gray]{#1}}%
      \expandafter\def\csname LTw\endcsname{\color{white}}%
      \expandafter\def\csname LTb\endcsname{\color{black}}%
      \expandafter\def\csname LTa\endcsname{\color{black}}%
      \expandafter\def\csname LT0\endcsname{\color{black}}%
      \expandafter\def\csname LT1\endcsname{\color{black}}%
      \expandafter\def\csname LT2\endcsname{\color{black}}%
      \expandafter\def\csname LT3\endcsname{\color{black}}%
      \expandafter\def\csname LT4\endcsname{\color{black}}%
      \expandafter\def\csname LT5\endcsname{\color{black}}%
      \expandafter\def\csname LT6\endcsname{\color{black}}%
      \expandafter\def\csname LT7\endcsname{\color{black}}%
      \expandafter\def\csname LT8\endcsname{\color{black}}%
    \fi
  \fi
  \setlength{\unitlength}{0.0500bp}%
  \begin{picture}(5102.00,4250.00)%
    \gplgaddtomacro\gplbacktext{%
      \csname LTb\endcsname%
      \put(540,640){\makebox(0,0)[r]{\strut{}0}}%
      \put(540,1483){\makebox(0,0)[r]{\strut{}5}}%
      \put(540,2325){\makebox(0,0)[r]{\strut{}10}}%
      \put(540,3168){\makebox(0,0)[r]{\strut{}15}}%
      \put(540,4010){\makebox(0,0)[r]{\strut{}20}}%
      \put(660,440){\makebox(0,0){\strut{}$0$}}%
      \put(1339,440){\makebox(0,0){\strut{}$2$}}%
      \put(2018,440){\makebox(0,0){\strut{}$4$}}%
      \put(2697,440){\makebox(0,0){\strut{}$6$}}%
      \put(3376,440){\makebox(0,0){\strut{}$8$}}%
      \put(4055,440){\makebox(0,0){\strut{}$10$}}%
      \put(4734,440){\makebox(0,0){\strut{}$12$}}%
      \put(80,2325){\rotatebox{90}{\makebox(0,0){\strut{}\strfact}}}%
      \put(2731,140){\makebox(0,0){\strut{}\invtemp}}%
    }%
    \gplgaddtomacro\gplfronttext{%
      \csname LTb\endcsname%
      \put(1203,3847){\makebox(0,0)[l]{\strut{}$\lattsz=24$}}%
      \csname LTb\endcsname%
      \put(1203,3647){\makebox(0,0)[l]{\strut{}$\lattsz=20$}}%
      \csname LTb\endcsname%
      \put(1203,3447){\makebox(0,0)[l]{\strut{}$\lattsz=16$}}%
      \csname LTb\endcsname%
      \put(1203,3247){\makebox(0,0)[l]{\strut{}$\lattsz=12$}}%
    }%
    \gplbacktext
    \put(0,0){\includegraphics{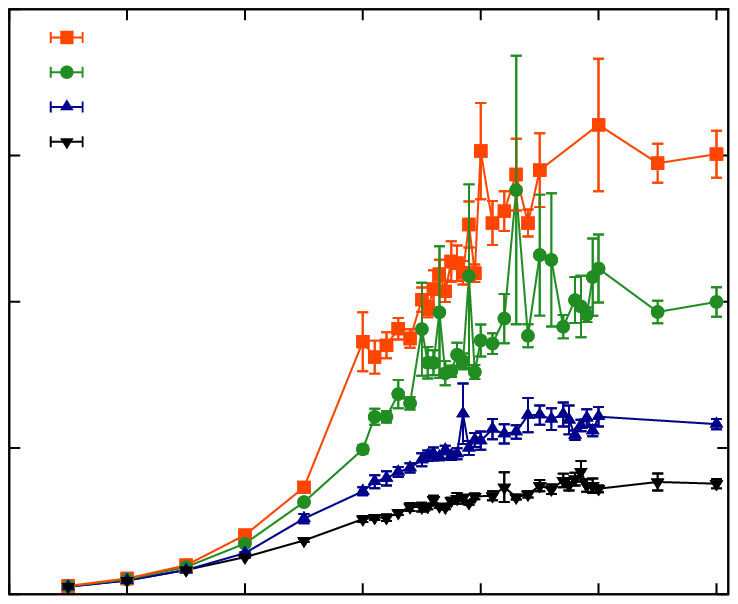}}%
    \gplfronttext
  \end{picture}%
\endgroup

%% file: etafit_struct_off.tex
\begingroup
  \makeatletter
  \providecommand\color[2][]{%
    \GenericError{(gnuplot) \space\space\space\@spaces}{%
      Package color not loaded in conjunction with
      terminal option `colourtext'%
    }{See the gnuplot documentation for explanation.%
    }{Either use 'blacktext' in gnuplot or load the package
      color.sty in LaTeX.}%
    \renewcommand\color[2][]{}%
  }%
  \providecommand\includegraphics[2][]{%
    \GenericError{(gnuplot) \space\space\space\@spaces}{%
      Package graphicx or graphics not loaded%
    }{See the gnuplot documentation for explanation.%
    }{The gnuplot epslatex terminal needs graphicx.sty or graphics.sty.}%
    \renewcommand\includegraphics[2][]{}%
  }%
  \providecommand\rotatebox[2]{#2}%
  \@ifundefined{ifGPcolor}{%
    \newif\ifGPcolor
    \GPcolortrue
  }{}%
  \@ifundefined{ifGPblacktext}{%
    \newif\ifGPblacktext
    \GPblacktexttrue
  }{}%
  \let\gplgaddtomacro\g@addto@macro
  \gdef\gplbacktext{}%
  \gdef\gplfronttext{}%
  \makeatother
  \ifGPblacktext
    \def\colorrgb#1{}%
    \def\colorgray#1{}%
  \else
    \ifGPcolor
      \def\colorrgb#1{\color[rgb]{#1}}%
      \def\colorgray#1{\color[gray]{#1}}%
      \expandafter\def\csname LTw\endcsname{\color{white}}%
      \expandafter\def\csname LTb\endcsname{\color{black}}%
      \expandafter\def\csname LTa\endcsname{\color{black}}%
      \expandafter\def\csname LT0\endcsname{\color[rgb]{1,0,0}}%
      \expandafter\def\csname LT1\endcsname{\color[rgb]{0,1,0}}%
      \expandafter\def\csname LT2\endcsname{\color[rgb]{0,0,1}}%
      \expandafter\def\csname LT3\endcsname{\color[rgb]{1,0,1}}%
      \expandafter\def\csname LT4\endcsname{\color[rgb]{0,1,1}}%
      \expandafter\def\csname LT5\endcsname{\color[rgb]{1,1,0}}%
      \expandafter\def\csname LT6\endcsname{\color[rgb]{0,0,0}}%
      \expandafter\def\csname LT7\endcsname{\color[rgb]{1,0.3,0}}%
      \expandafter\def\csname LT8\endcsname{\color[rgb]{0.5,0.5,0.5}}%
    \else
      \def\colorrgb#1{\color{black}}%
      \def\colorgray#1{\color[gray]{#1}}%
      \expandafter\def\csname LTw\endcsname{\color{white}}%
      \expandafter\def\csname LTb\endcsname{\color{black}}%
      \expandafter\def\csname LTa\endcsname{\color{black}}%
      \expandafter\def\csname LT0\endcsname{\color{black}}%
      \expandafter\def\csname LT1\endcsname{\color{black}}%
      \expandafter\def\csname LT2\endcsname{\color{black}}%
      \expandafter\def\csname LT3\endcsname{\color{black}}%
      \expandafter\def\csname LT4\endcsname{\color{black}}%
      \expandafter\def\csname LT5\endcsname{\color{black}}%
      \expandafter\def\csname LT6\endcsname{\color{black}}%
      \expandafter\def\csname LT7\endcsname{\color{black}}%
      \expandafter\def\csname LT8\endcsname{\color{black}}%
    \fi
  \fi
  \setlength{\unitlength}{0.0500bp}%
  \begin{picture}(5102.00,4250.00)%
    \gplgaddtomacro\gplbacktext{%
      \csname LTb\endcsname%
      \put(600,640){\makebox(0,0)[r]{\strut{}$10^{0}$}}%
      \put(600,1483){\makebox(0,0)[r]{\strut{}$10^{1}$}}%
      \put(600,2325){\makebox(0,0)[r]{\strut{}$10^{2}$}}%
      \put(600,3168){\makebox(0,0)[r]{\strut{}$10^{3}$}}%
      \put(600,4010){\makebox(0,0)[r]{\strut{}$10^{4}$}}%
      \put(950,440){\makebox(0,0){\strut{}$12$}}%
      \put(1782,440){\makebox(0,0){\strut{}$14$}}%
      \put(2502,440){\makebox(0,0){\strut{}$16$}}%
      \put(3138,440){\makebox(0,0){\strut{}$18$}}%
      \put(3707,440){\makebox(0,0){\strut{}$20$}}%
      \put(4221,440){\makebox(0,0){\strut{}$22$}}%
      \put(4691,440){\makebox(0,0){\strut{}$24$}}%
      \put(20,2325){\rotatebox{90}{\makebox(0,0){\strut{}\strfact [log, shifted]}}}%
      \put(2761,140){\makebox(0,0){\strut{}\lattsz [log]}}%
    }%
    \gplgaddtomacro\gplfronttext{%
    }%
    \gplbacktext
    \put(0,0){\includegraphics{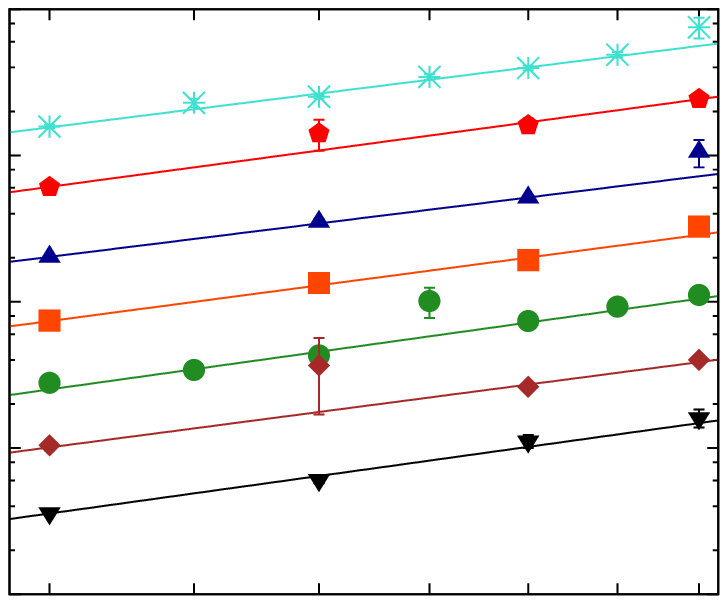}}%
    \gplfronttext
  \end{picture}%
\endgroup

%% file: negutrap.bbl
\begin{thebibliography}{80}
\bibitem{footnote1} While all optical lattice experiments involve
  inhomogeneity from the confining potential, in some, randomness,
  e.g.~in the form of laser speckle is also being deliberately
  introduced to study the interplay of disorder and interactions.
\bibitem{footnote2} In solids, the hopping parameter $t$ can be
  influenced through the application of pressure.
\end{thebibliography}
